# Low Threshold Parametric Decay Instabilities in ECRH experiments at toroidal devices

E Z Gusakov, A Yu Popov

Ioffe Institute, St. Petersburg, Russia

e-mail: Evgeniy.Gusakov@mail.ioffe.ru

**Abstract.** The experimental conditions leading to substantial reduction of backscattering decay instability threshold in ECRH experiments in toroidal devices are analyzed. It is shown that drastic decrease of threshold is provided by non monotonic behaviour of plasma density, which is often observed due to so-called density-pump-out effect or presence of magnetic islands, and by poloidal magnetic field inhomogeniety making possible localization of ion Bernstein decay waves. The corresponding ion Bernstein wave gain and the parametric decay instability pump power threshold is calculated. The possible experimental consequences of easy backscattering decay instability excitation are discussed.

## 1. Introduction

Electron cyclotron resonance heating (ECRH) at power level of up to 1 MW in a single microwave beam is often used in present day tokamak and stellarator experiments and planed for application in ITER for neoclassical tearing mode island control. Parametric decay instabilities (PDI) leading to anomalous reflection or absorption of microwave power are believed to be deeply suppressed in tokamak MW power level ECR O-mode and second harmonic X-mode heating experiments utilizing gyrotrons [1-3]. According to theoretical analysis of PDI thresholds [1-3], the typical RF power at which these nonlinear effects can be excited at tokamak plasma parameters is around 1 GW, which is only possible with free electron laser application. The physical reason for such a deep suppression is provided by strong convective losses of the daughter waves from the decay region either in the plasma inhomogeniety direction or along the magnetic field [1-3]. In the first case the daughter waves amplification in the narrow region, where the decay condition $\Delta k = k_1 - k_2 - k_0 = 0$, ($k_1$, $k_2$ and $k_0$ - wave numbers of two daughter waves and pump, correspondingly), is fulfilled in inhomogeneous plasma, is described by the so called Piliya – Rosenbluth coefficient [4-6]

$$S_{PR} = \exp\left\{\frac{\pi \gamma_0^2 \ell^2}{|v_1 v_2|}\right\} \qquad (1)$$

where $\gamma_0$ - is the maximal PDI growth rate in homogeneous plasma, proportional to the pump wave amplitude, $v_1$ and $v_2$ - the daughter wave group velocities and $\ell^2 = \left|d(\Delta k)/dx\right|^{-1/2}$. As it is clear from this formula the PDI threshold can be lowered by the growth of the pump field or/and by decrease of the daughter wave group velocity. Both effects occur in the case of induced backscattering in the vicinity of the pump wave upper hybrid resonance (UHR) [7] explaining easy PDI excitation in EBW heating experiments. Until recently it was the only situation where backscattering PDI was observed in toroidal devices at 100 kW ECRH power level [8]. This mechanism of the PDI threshold lowering is however not applicable to the first harmonic O-mode and second harmonic X-mode heating experiments where no UHR exists for the pump wave. Therefore it is taken for granted that wave propagation and absorption in these experiments is well described by linear theory and thus predictable in detail.

However during the last decade a "critical mass" of observations has been obtained evidencing presence of anomalous phenomena in ECRH experiments at toroidal devices. First of all, non local electron transport was shown to accompany ECRH in some devices indicating that the RF power is not deposited in the regions predicted by standard theory, but is rather redistributed very quickly all over the plasma [9]. Secondly, fast ion generation and ion heating was observed during ECRH pulse under conditions when energy exchange between electrons and ions should be very low [10]. And finally, last year the first observations of the backscattering signal in the 200 – 600 kW level second harmonic ECRH experiment at Textor tokamak were reported [11, 12]. This signal down shifted in frequency by approximately 1 GHz, which is close to the lower hybrid or ion Bernstein (IB) wave frequency under the Textor conditions, was strongly modulated in amplitude at the m=2 magnetic island frequency. This observation performed at the modest RF power under conditions when no pump wave UHR was present provides an indication that probably a low threshold PDI excitation is possible in ECRH experiment at certain conditions which are somehow associated with the presence of a magnetic island in tokamak discharge.

The novel low threshold mechanism of the PDI excitation was proposed in [13, 14] based on analysis of the actual Textor density profile. It was shown that the local maximum of the plasma density which is usually observed in the O-point of magnetic island at Textor [15, 16] can lead to localisation of the low frequency ion Bernstein (IB) decay wave and thus to suppression of IB wave convective losses in radial direction. A more complicated 2D analysis of the IB wave propagation accounting for the poloidal inhomogeniety of magnetic field in toroidal plasma have shown possibility of IB wave localisation in the poloidal direction as well [17]. The threshold of the strict backscattering PDI was calculated in this case and shown to be more than four orders of magnitude lower than predictions of standard theory (in the range of 50 kW for the Textor experiment parameters).

In the present paper we give a more detailed description of the analysis of backscattering PDI accompanying the second harmonic ECRH experiment in toroidal device in the case of local plasma density maximum. The treatment is not limited to the case of strict backscattering. The IB wave amplification due to the PDI is calculated and corresponding thresholds are evaluated. The experimental conditions under which the proposed mechanism can be of importance are discussed.

## 2. The basic equations

To elucidate the physics of the PDI amplification we use the most simple but nevertheless relevant to the experiment [12] Cartesian co-ordinate system $(x, y, z)$ with its origin located at the mid-plane of the torus, the $x-$ axis being opposite to the major radius $R$ and the $y, z$ co-ordinates imitating poloidal and toroidal directions, respectively. For the sake of simplicity we assume the pump frequency exceeding both the electron cyclotron and plasma frequency so that the following strong inequality holds: $\omega_i^2 \gg \omega_{pe}^2, \omega_{ce}^2$, which is the case in the TEXTOR electron cyclotron resonance heating experiments. We neglect also a weak dependence of the high frequency wave numbers on coordinate that allows us to introduce the pump wave in the form $E_{iy} = a_i(y,z)\exp(ik_{ix}x - i\omega_i t)$ describing a wide microwave beam propagating from the launching antenna inwards plasma in the tokamak (stellarator) mid-plane with amplitude $a_i = [8\pi P_i / (\pi\rho^2 c)]^{1/2} \exp[-(y^2 + z^2)/2\rho^2]$, where $P_i$ is the pump wave power and $\rho$ is the beam radii. The almost backscattered X-mode wave is introduced as $E_{sy} = a_s(\vec{r})\exp(-ik_{sx}x - ik_{sy}y - i\omega_s t)$, $k_{sx} \gg k_{sy}$. The basic equations describing the backscattered wave generation and its convective losses from the decay region as well as the low-frequency, $\Omega = \omega_i - \omega_s \ll \omega_i$, electrostatic IB wave $\vec{E} = -\vec{\nabla}\varphi \exp(i\Omega t)$ are as follows:

$$\left(\frac{\partial^2}{\partial x^2} + k_{sx}^2\right) E_{sy} = i\frac{4\pi\omega_s}{c^2} j_{sy} \tag{2}$$

$$\int_{-\infty}^{\infty} d\vec{r}' \hat{D}\left(\vec{r} - \vec{r}', (\vec{r} + \vec{r}')/2\right) \varphi(\vec{r}') = 4\pi\rho_\Omega \tag{3}$$

The nonlinear current in (2) $j_{sy} = e\delta n_\Omega u_{iy} \simeq -\frac{e}{m_e c^2}\frac{\omega_{pe}^2}{\omega_{ce}^2}\frac{\partial^2\varphi}{\partial x^2} E_{iy}$ is given by a product of an electron density perturbation $\delta n_\Omega$ produced by a low-frequency small scale decay wave and the quiver electron velocity $u_{iy}$ associated with the pump wave. The nonlinear charge density

$\rho_\Omega$ in (3), generated by the ponderomotive force, is responsible for coupling of low and high frequency waves. In the LH frequency range it takes a form $\rho_\Omega = \frac{e}{4\pi m_e} \frac{\omega_{pe}^2}{\omega_{ce}^2 \omega_i^2} \frac{\partial}{\partial x}\left[ E_{iy}^* \partial \frac{E_{sy}}{\partial x} + E_{sy} \frac{\partial E_{iy}^*}{\partial x} \right]$. In weakly inhomogeneous plasma the operator $\hat{D}$ in the integral equation (3), exhibiting much stronger dependence on the first argument $\vec{r} - \vec{r}'$ than on the second - $(\vec{r}+\vec{r}')/2$, associated with the plasma inhomogeneity, can be represented as

$$\hat{D}\left[\vec{r}-\vec{r}', \frac{\vec{r}+\vec{r}'}{2}\right] = \frac{1}{(2\pi)^3} \int D\left[\vec{q}, \frac{\vec{r}+\vec{r}'}{2}\right] \exp[i\vec{q}(\vec{r}-\vec{r}')] d\vec{q},$$

where

$$D\left(\vec{q}, \frac{\vec{r}+\vec{r}'}{2}\right) = q^2\left(1 + \chi_e\left(\vec{q}, \frac{\vec{r}+\vec{r}'}{2}\right) + \chi_i\left(\vec{q}, \frac{\vec{r}+\vec{r}'}{2}\right)\right) \qquad (4)$$

(The integrals throughout the paper, if it is not determined in another way explicitly, are evaluated over the interval $[-\infty, \infty]$.) The electron susceptibility $\chi_e$, entering expression (4) is provided by cold homogeneous plasma model in the form $\chi_e = \frac{q_\perp^2}{q^2}\frac{\omega_{pe}^2}{\omega_{ce}^2} - \frac{q_\parallel^2}{q^2}\frac{\omega_{pe}^2}{\Omega^2}$, where wave vector components along and across magnetic field are given correspondingly by $q_\parallel = q_z \cos\phi + q_y \sin\phi$, $q_\perp = (q_x^2 + (q_y\cos\phi - q_z\sin\phi)^2)^{1/2}$, where $\tan\phi = B_{pol}/B_{tor}$. For the ion susceptibility $\chi_i$ we use the representation derived in [18], which is valid for $\omega > 2\omega_{ci}$

$$\chi_i = \frac{2\omega_{pi}^2}{q^2 v_{ti}^2}\left[1 - X\left(\frac{\Omega}{q_\perp v_{ti}}\right) - \left(\cot\left(\frac{\pi\Omega}{\omega_{ci}}\right) + \frac{i}{\sqrt{\pi}}\frac{\omega_{ci}}{|q_\parallel|v_{ti}}\sum_{m=-\infty}^{\infty}\exp\left(-\frac{(\Omega-m\omega_{ci})^2}{q_\parallel^2 v_{ti}^2}\right)\right)Y\left(\frac{\Omega}{q_\perp v_{ti}}\right)\right],$$

where

$$X(\xi) + iY(\xi) = \frac{\xi}{\sqrt{\pi}}\int_{-\infty}^{\infty}\frac{\exp(-t^2)}{t-\xi-io}dt.$$

3. **The simplified model.**

Taking into account that the PDI amplification is enhanced when the daughter wave group velocity decreases [4-6] we shall consider solutions of (2), (3) in the vicinity of the low frequency wave turning point in $x$ direction ($x = x_0$). Here conditions $D|_{q_{x0},\Omega_0,x_0} \equiv D|_0 = 0$, $\partial D/\partial q_x|_0 = 0$ hold for wave at frequency $\Omega = \Omega_0$, wave numbers $q_x = q_{x0}$, $q_y = 0$ and its group velocity goes to zero. Due to periodicity of $\cot(\pi\Omega/\omega_{ci})$ function solution of equations

$D|_0 = 0$ and $\partial D/\partial q_x|_0 = 0$ is not unique and, as it was shown in [14, 17], it is easy to find one approximately satisfying the pump Bragg backscattering condition ($q_{x0} \simeq k_{ix} + k_{sx}$). Moreover, in attempt to explain the extremely low pump power level at which backscattering correlated to magnetic activity was observed in [11, 12] we shall, following [13, 14], make use of non monotonous profile of plasma density often observed by diagnostics possessing high temporal and spatial resolution in the vicinity of magnetic islands [15, 16, 19]. Namely, we assume, as it is shown in experiment [16], that a local density maximum is associated with the island O - point and therefore conditions $\partial D/\partial x|_0 = 0$ and $\partial^2 D/\partial x^2|_0 > 0$ hold in its vicinity (been slightly shifted from the density maximum due to magnetic field inhomogeniety). In these circumstances two nearby turning points ("warm" to "hot" mode) can exist in plasma for high harmonic IB wave and it can be trapped in $x$ direction, if additional condition $\partial^2 D/\partial q_x^2|_0 > 0$ holds. The corresponding density profile and IB wave dispersion curve $q_x(x)$ calculated for frequency $\Omega_0 = 0.86$ GHz, ion temperature $T_i = 500$ eV is shown in figure 1. The described possibility of IB wave localization in a plasma waveguide provides an argument in favor of the PDI threshold lowering in the case of the waveguide eigen mode excitation. The physical reason for it is related to the suppression of convective wave energy losses in the inhomogeniety direction. It is important to note that due to magnetic field dependence on major radius IB wave trapping is possible also in the poloidal direction. It occurs when IB wave is excited with small parallel wave number close to the equatorial plane at the low magnetic field side of the torus. To illustrate the above conclusions we treat the dispersion relation for the IB wave $D = 0$, where $D$ is given by equation (4), using ray tracing procedure for Textor experimental conditions [15] accounting for dependence of the magnetic field on two coordinates $\omega_{ci}(x,y) = \omega_{ci}(R_0) \cdot R_0/R(x,y)$ with $R_0$ and $R$ being major radii at the magnetic axis and in the actual point and taking $q_{x0}(\partial^2 D/\partial x^2|_0)^{-1/2} \approx 5$ cm. We neglect here small perturbations of the magnetic field associated with the island and keep in our simulation only the equilibrium magnetic field. Then, we find numerically a solution of the set of Hamiltonian equations

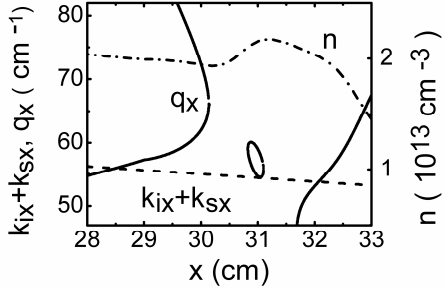

Figure 1. The IB wave dispersion curves (solid line) and plasma density (dash dotted line) in the magnetic island vicinity. Dashed line - $k_{ix}+k_{sx}$, $\Omega_0 = 0.86$ GHz, $T_i = 500$ eV.

$$\frac{\partial \vec{q}}{\partial s} = \frac{\partial D(\vec{q},\vec{r})}{\partial \vec{r}}, \qquad \frac{\partial \vec{r}}{\partial s} = -\frac{\partial D(\vec{q},\vec{r})}{\partial \vec{q}} \qquad (5)$$

describing the ray behavior in the 4D phase space $\vec{r} = (x,y); \vec{q} = (q_x, q_y)$ in a vicinity of the turning point of the IBW located near the O-point of the magnetic island. We introduce above $ds = dt \partial D / \partial \Omega \big|_0^{-1}$, where $s$ is an effective IBW trajectory length. Evaluating (5) for different initial poloidal and radial wave numbers with the starting point of simulation at $\vec{r}_s = (0,0)$ results in dependencies of trajectory coordinates on its "length" $y(s)$, $x(s)$ and in phase portraits $q_x(x)$ and $q_y = q_y(y)$ shown in figures 2 - 4. As it is seen in figure 2 and figure 3, the phase portraits justify the finite behavior of the 4-D ray trajectory in both $x$ and $y$ direction, and therefore proves the 2D trapping of the ray and, as a result, of the IB wave. IB waves possessing small enough poloidal wave number are localized close to equatorial plane and do not cross the high harmonic ion cyclotron resonance layer, thus not suffering from heavy damping. The corresponding oscillatory trajectory behavior is presented in figure 4. Upon making Taylor expansion of the dispersion relation $D=0$ at $\vec{q}_0 = (q_{x0}, 0)$ and $\vec{r}_0 = (x_0, 0)$ up to the first non-vanishing terms we reduce the integral equation (3) to

$$\left\{ \Delta D - D_{qq} \frac{\partial^2}{\partial x^2} - 2D_{xq}(x-x_0)\frac{i\partial}{\partial x} - iD_{xq} + \eta_0 \sin^2\phi \frac{\partial^2}{\partial y^2} + D_{xx}(x-x_0)^2 + D_{yy} y^2 - \delta D \right\} b = \\ 4\pi \rho_\Omega \exp\left[ i q_{\perp 0} x + i q_z (z - y \cot\phi) \right] \qquad (6)$$

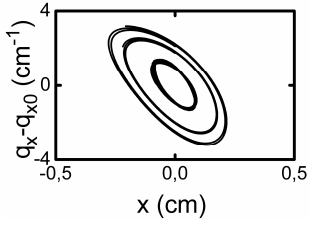

Figure 2. The phase portrait $q_x(x) - q_{x0}$, $q_{x0} = 59.7 cm^{-1}$, $\Omega_0 = 0.86$ GHz, $T_i = 500$ eV

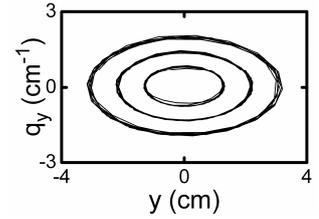

Figure 3. The phase portrait $q_y(y)$, $\Omega_0 = 0.86$ GHz, $T_i = 500$ eV.

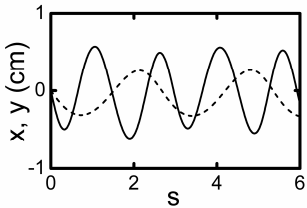

Figure 4. Radial (dashed curve) and poloidal (solid curve) ray position versus trajectory "length".

where we assume $\cos\phi \approx 1$ and introduce new notations

$$\delta\Omega = \Omega - \Omega_0, \quad \Delta D = \delta\Omega \frac{\partial D}{\partial \Omega}\bigg|_0, \quad \eta_0 = \frac{\omega_{pe}^2}{\Omega^2}\bigg|_0, \quad D_{qq} = \frac{\partial^2 D}{2 \partial q_x^2}\bigg|_0,$$

$$D_{xq} = \frac{\partial^2 D}{2 \partial q_x \partial x}\bigg|_0, \quad D_{xx} = \frac{\partial^2 D}{2 \partial x^2}\bigg|_0, \quad D_{yy} = \frac{\partial^2 D}{2 \partial y^2}\bigg|_0 \equiv -\frac{q_{\perp 0}^2}{L_y^2},$$

$$L_y^{-2} = \frac{\omega_{pi}^2(x)}{q_x^2 v_{ti}^2} \csc^2\left(\pi \frac{\Omega}{\omega_{ci}}\right) Y\left(\frac{\Omega}{q_x v_{ti}}\right) \frac{\pi \Omega}{\omega_{ci}} \frac{1}{rR}\bigg|_0.$$ The amplitude

$b = \varphi \exp\left[ i q_{\perp 0} x + i q_z (z - y \cdot \cot\phi) \right]$ in (6) corresponds to the IB wave propagating almost perpendicular to magnetic field and the perturbation term

$$\delta D = \frac{q_z^2}{4\sin^2\phi}\left[\frac{q_{x0}D''' - 4D_{qq}}{q_{x0}^2}\frac{\partial^2}{\partial x^2} + \frac{6D_{qq}}{q_{x0}^2}\frac{\partial^2}{\partial y^2}\right] - i\frac{2}{\sqrt{\pi}}\frac{\omega_{pi}^2}{v_{ti}^2}\frac{\omega_{ci}}{|q_\parallel|v_{ti}}\sum_{m=-\infty}^{\infty}\exp\left(-\frac{(\Omega - m\omega_{ci})^2}{q_\parallel^2 v_{ti}^2}\right)Y\left(\frac{\Omega}{q_{\perp 0}v_{ti}}\right), \quad (7)$$

is describing the diffractive losses in the toroidal direction (the first term) and IBW damping (the last one). We introduce also the notation: $q_{x0}D''' = q_{x0}\partial^3 D/\partial q_\perp^3 |_0$. In this consideration we neglect weak poloidal and moreover very weak toroidal density inhomogeniety in the magnetic island and therefore suppose wave number $q_z$ constant.

**The PDI analyses and discussion**

Assuming the IB wave damping, PDI pumping and convection in toroidal direction weak we account for them using the perturbation theory approach [20]. In the zero order approximation we neglect $\delta D$ and $\rho_\Omega$ in (6) and obtain equation which can be solved by separation of variables. The corresponding expression for the IB eigen mode trapped in radial and poloidal direction and possessing $q_z = 0$, which describes propagation without convective losses in toroidal direction, is given by

$$b(x,y) = \exp\left(-i\frac{D_{xq}}{2D_{qq}}(x-x_0)^2\right)\varphi_k(x)\varphi_l(y) = \quad (8)$$

$$H_k\left(\frac{x-x_0}{\delta_x}\right)H_l\left(\frac{y}{\delta_y}\right)\exp\left[-i\frac{D_{xq}}{2D_{qq}}(x-x_0)^2 - \left(\frac{x-x_0}{\sqrt{2}\delta_x}\right)^2 - \left(\frac{y}{\sqrt{2}\delta_y}\right)^2\right]$$

where $H_k$ is standing for Hermitian polynomial, the size of the IB mode localization region is $\delta_x = (L_x/q_{x0})^{1/2}(D_{qq})^{1/4}$, $\delta_y = \sin\phi(L_y/q_{x0})^{1/2}(\omega_{pe}/\Omega_0)^{1/2}$, where we introduce $2q_{x0}^2 L_x^{-2} \equiv \left[\partial^2 D/\partial x^2 - (\partial^2 D/\partial x \partial q_x)^2 / (\partial^2 D/\partial q_x^2)\right]_0$ which is supposed to be positive, as it is in the Textor experiment. The exact value of the mode frequency is determined by following quantization condition

$$\delta\Omega_{k,l} = \partial D/\partial\Omega\Big|_0^{-1}\left[\sqrt{D_{qq}}\frac{q_{x0}}{L_x}(2k+1) - \sin\phi\frac{\omega_{pe}}{\Omega_0}\frac{q_{x0}}{L_b}(2l+1)\right].$$

At the next step of the perturbation analysis procedure we account for IB wave damping, PDI pumping and convection in toroidal direction. Expressing the BS wave amplitude from (2) in terms of the IB wave potential and introducing $\Delta K = q_{x0} - k_{ix} - k_{sx}$ we obtain the nonlinear charge density in the form

$$\rho_\Omega\left[b(x,y)\right]\exp\left(iq_{x0}x - iq_z(y\cot\phi - z)\right) = -i\frac{q_{x0}^3}{16\pi}\frac{\omega_{pe}^4}{\omega_i^2\omega_{ce}^2}\frac{|a_i|^2}{H^2}\exp\left(-\frac{y^2+z^2}{\rho^2} + ik_{sy}y\right)\int_{-\infty}^{x}dx'\exp(i\Delta K(x-x'))b(x',y).$$

Substituting the zero order solution (8) into this expression and in $\delta D$ and requiring no variation of the eigen frequency $\delta \Omega_{kl}$ with the accuracy up to the first order we obtain equation for the toroidal wave number determining damping or PDI amplification of the IB wave

$$\int_{-\infty}^{\infty}\int_{-\infty}^{\infty} \varphi_k(x)\varphi_l(y)\{\delta D[\varphi_k(x)\varphi_l(y)] - 4\pi\rho_\Omega[\varphi_k(x)\varphi_l(y)]\exp(iq_{x0}x - iq_z(y\cot\phi - z))\} dy dx = 0 \quad (9)$$

The necessary condition for the PDI onset is provided by excess of the pump contribution to $q_z$ over the damping contribution. When the ion cyclotron harmonic is far from the IB wave trapping region ($(\Omega - m\omega_{ci})/q_z v_{ti} \gg 1$) the damping is negligible and the imaginary part of toroidal wave number $q_z''$ is given by expression

$$q_z'' = |\sin\phi| q_{x0}^{5/2} \delta_x^{3/2} \frac{\sqrt{2}}{2} \frac{\omega_{pe}^2}{\omega_i \omega_{ce}} \frac{a_i}{H} \alpha_{k,l}, \quad (10)$$

where

$$\alpha_{k,l}^2 = \frac{2}{\pi k! 2^k l! 2^l} \frac{\delta_y^{-1} \int_{-\infty}^{\infty} dy [\exp(-y^2/\rho^2 + ik_{sy}y)\varphi_l^2(y)]}{(2k+1)\left(1 + \frac{D_{qx}^2}{D_{qq}^2}\delta_x^4\right)(q_{x0}D''' - 4D_{qq}) + 6D_{qq}(2l+1)\frac{\delta_x^2}{\delta_y^2}} \times$$

$$\delta_x^{-2} \int_{-\infty}^{\infty} dx \varphi_k(x) \exp\left(i\frac{D_{xq}}{2D_{qq}}x^2\right) \int_{-\infty}^{x} dx' \exp(i\Delta K(x-x'))\varphi_k(x')\exp\left(-i\frac{D_{xq}}{2D_{qq}}x'^2\right).$$

For the fundamental mode $k = 0$ and $l = 0$ we get

$$\alpha_{0,0}^2 \simeq \frac{2\sqrt{\pi}\delta_y^2}{\left[(1 + D_{qx}^2 \delta_x^4/D_{qq}^2)(q_{x0}D''' - 4D_{qq})\delta_y^2 + 6D_{qq}\delta_x^2\right]} \frac{\rho}{\sqrt{\rho^2 + \delta_y^2}} \times$$

$$\frac{D_{qq}}{\sqrt{D_{qq}^2 + D_{xq}^2\delta_x^4}} \exp\left(-\frac{\Delta K^2 \delta_x^2 D_{qq}^2}{D_{qq}^2 + D_{xq}^2\delta_x^4} - \frac{k_{sy}^2 \delta_y^2 \rho^2}{4(\delta_y^2 + \rho^2)}\right).$$

The exponential term in the later expression is related to mismatch of decay conditions in radial and poloidal direction leading to suppression of nonlinear three-wave coupling. In the case of wide pump beam $\rho > \sqrt{2l+1}\delta_y \cot\phi$ the sufficient condition for the PDI onset is given by expression $\Gamma = 2\sqrt{2\pi}\rho q_z'' > 1$ determining large enough gain of the IB wave when propagating across pump beam along toroidal direction. In the opposite case at $\rho < \sqrt{2l+1}\delta_y \cot\phi$ the poloidal gain over the IB mode localization region provided by exponential factor $\exp[iq_z y\cot\phi]$ dominates and the PDI threshold takes a form

$\Gamma = 2\sqrt{2l+1}\delta_y \cot\phi q_z'' > 1$. Dependence of this gain on the IB radial mode number, computed accounting for the IB wave damping, is shown in figure 5 for case ($\vartheta \equiv \arctan(k_{sy}/k_{sx}) = 2°$). Computations are performed for poloidal IB mode ($l=0$), TEXTOR experiment parameters and different plasma densities. As it is seen, because of radial decay condition mismatch ($\Delta K \neq 0$) the gain is not always maximal for the fundamental radial mode. Nevertheless a substantial gain is predicted by theory for the routinely used ECRH power of 400 kW. As it is shown in figure 6, the backscattered wave frequency corresponding to the maximal gain is decreasing with the plasma density increase. This dependence is consistent with corresponding dependence of the back scattered wave frequency shift observed in [12].

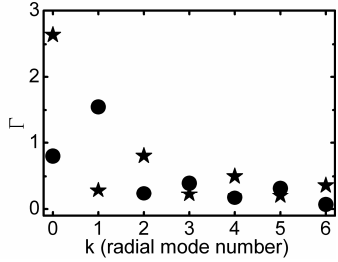

Figure 5. Dependence of the IB wave gain on the radial mode number at $l = 0$; $P = 400$ kW, $T_i = 500$ eV, $\vartheta = 2°$; Circles – $n = 2\cdot 10^{13}$ cm$^{-3}$, $\Omega_0/2\pi = 0.832$ GHz; Stars - $n = 3\cdot 10^{13}$ cm$^{-3}$, $\Omega_0/2\pi = 0.923$ GHz.

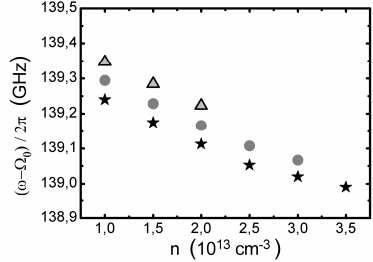

Figure 6. Dependence of the frequency of the scattered wave corresponding to maximal IB wave gain on plasma density. (Triangles - $T_i = 300$ eV; Circles - $T_i = 600$ eV; Stars - $T_i = 900$ eV).

Dependence of the gain on poloidal scattering angle calculated for IB mode numbers $k=0, l=0$ (dashed curve) and $k=0, l=1$ (solid curve) is shown in figure 7 for Textor parameters. As it is seen there substantial amplification of trapped IB wave is foreseen for scattering angles in the range $|\vartheta| \leq 10°$. It should be underlined that at sufficiently high amplification value it can lead to excitation of absolute PDI which grows slower, but saturates at much higher level due to nonlinear effects. The later instability can lead to strong anomalous reflection and, as a result, to fast variation of heating power deposition profile. The PDI power threshold provided by condition $\Gamma > 1$ is given by formula

$$P_{th} = \frac{cH^2}{2\pi\alpha_{k,l}^2} \frac{\omega_i^2 \omega_{ce}^2}{\omega_{pe}^4} \left(\frac{\rho^2}{(2l+1)\delta_y^2} \frac{1}{q_x^5 \delta_x^3}\right)\bigg|_0 \tag{10}$$

In the case of typical TEXTOR experimental parameters ($H = 19$ kGs, $f_i = 140$ GHz, $n = 3\times 10^{13}$ cm$^{-3}$, $T_i = 500$ eV, $\rho = 1$ cm) assuming for magnetic island density variation $\delta n/n = 0.11$ and width $w = 2$ cm, as measured in [16], we obtain for the fundamental IB mode ($k=0, l=0$) and $\vartheta = 0°$ the threshold value $P_{th} \approx 50$ kW which is overcome in the experiment. The corresponding frequency shift $\Omega_0 = 0.923$ GHz is close to that observed in the experiment [12], $\delta\Omega_{0,0}/(2\pi) = 5.7$ MHz, $\delta_y = 1.1$ cm, $\delta_x = 0.6$ cm and $\Delta K \delta_x \ll 1$. It is important to stress that the backscattering PDI threshold is overcome at typical gyrotron power range not only in medium scale devices, but also in reactor scale machines at fusion conditions. For example, Figure 8 shows the PDI gain dependence on scattering angle calculated for ITER conditions (P = 1 MW, $H = 50$ kGs, $f_i = 170$ GHz, $n = 10^{14}$ cm$^{-3}$, $T_i = 5$ keV, $\rho = 5$ cm) assuming for magnetic island density variation $\delta n/n = 0.1$ and width $w = 10$ cm. As it is seen, very high gain close to $10^5$ is expected there which may lead to substantial anomalous reflection of heating power already at the convective instability stage without excitation of absolute PDI.

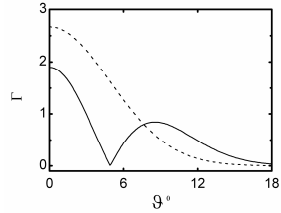

Figure 7. Dependence of the IB wave gain on the poloidal scattering angle for Textor parameters; P = 400 kW, T$_i$ = 500eV, n = 3·10$^{13}$ cm$^{-3}$, $\Omega_0/2\pi$ = 0.923 GHz. Dashed curve – k = 0, l = 0; solid curve – k = 0, l = 1.

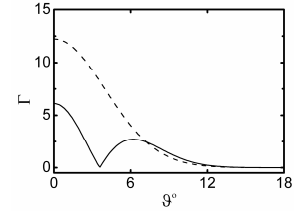

Figure 8. Dependence of the IB wave gain on the poloidal scattering angle for ITER parameters; P = 1 MW, T$_i$ = 5 keV, n = 10$^{14}$ cm$^{-3}$, $\Omega_0/2\pi$ = 2.31 GHz. Dashed curve – k = 0, l = 0; solid curve – k = 0, l = 1.

## 4. Conclusions

The drastic, compared to predictions of the standard theory [1-3], decrease of the PDI threshold shown in this paper is explained by complete suppression of IB wave radial and poloidal convective losses and their substantial reduction in the third direction. This mechanism is based first of all on non-monotonous dependence of plasma density on radial coordinate. Such a dependence observed in tokamaks in the presence of magnetic islands [15, 16] is associated with their magnetic confinement properties which, we believe, are not specific for the TEXTOR experimental conditions (see for instance [19]) and may lead to easy PDI excitation in ECRH experiments in other toroidal devices where magnetic islands usually exist. In the absence of magnetic islands the hollow density profiles are also often observed during ECRH experiments, in particular in stellarators, where they are explained by a density pump out effect [21]. Moreover it should be mentioned that not only magnetic islands, but also blobs and drift wave density perturbations, which are as well elongated along magnetic field, in the case of intensive enough turbulence can result in IB wave trapping. Similar effect leading to reduction of PDI threshold can occur also on the plasma discharge axis.

Another important feature of the low threshold PDI onset mechanism is related to the poloidal inhomogeniety of magnetic field in toroidal devices. It provides the possibility of IB wave localization in poloidal direction at the low magnetic field side of the torus (in the equatorial plane in the tokamak case).

Backscattering PDI can lead to reduction of ECRH efficiency and extremely fast change of its localization providing an alternative explanation for the so called "nonlocal electron transport effect" [9]. Absorption of parametrically driven IB wave can be responsible for ion acceleration often observed in ECRH experiments (see [17] and references there).

It should be stressed that backscattering PDI is potentially dangerous for the ECRH neoclassical magnetic island control method planned for application at ITER. In this relation understanding of physical mechanism of density peaking in the island and development of methods of its control is very important on one hand. On the other hand methods of control and suppression of backscattering PDI which were developed in small scale model experiments [22, 23] deserve more attentive study and testing in large scale ECRH experiments.

## 5. Acknowledgments

Financial support by RFBR grants 10-02-90003-Bel, 10-02-00887, NWO-RFBR Centre of Excellence on Fusion Physics and Technology (grant 047.018.002) and scientific school grant- 6214.2010.2 is acknowledged.